# Broadband Microwave Spectroscopy for Two-Dimensional Material Systems

Antonio L. Levy* and Neil M. Zimmerman
*Physical Measurement Laboratory, National Institute of Standards and Technology (NIST), Gaithersburg, Maryland, 20899-8171, USA*

## I. Abstract

In recent years, interesting materials have emerged which are only available as µm-scale flakes, and whose novel physics might be better understood through broadband microwave spectroscopy; examples include twisted bilayer graphene [1], 2D materials in which many-body phases are observed [2], and artificial lattices for analog quantum simulations [3]. Most previous techniques are unfortunately not sensitive for flakes below mm lateral sizes. We propose a simple technique which does not require sophisticated sample preparation nor Ohmic contact and show through theory and simulations that one will be able to qualitatively measure spectral features of interest, and quantitatively measure the frequency-dependent complex conductivity.

## II. Introduction

In recent years, two-dimensional materials, such as twisted bilayer graphene [1], 2D materials in which many-body phases are observed [2], and artificial lattices [4, 3], have emerged; the novel physics of these materials might be better understood through broadband microwave spectroscopy [5, 6, 7]. Also, understanding the physics behind the novel phenomena exhibited by these material systems may prove necessary to realizing applications based on them. One difficulty with such measurements is that often the thin films can only be grown in small flakes, of order 1 µm in lateral size; this work proposes a new broadband measurement which can be conveniently applied to such small flakes and does not require Ohmic contact.

Broadband conductivity spectroscopy has helped demystify the physics of some thin film materials. Such a system's conductivity spectrum depends strongly on its energy level spectrum, and thus is a direct reflection of the ground state and excited state physics. Previously, broadband microwave spectroscopy has been used to study novel physics of thin film materials which can be grown in large mm-scale flakes, such as charge density wave ground states of 2D electron systems [5, 8, 9, 10, 11, 12, 13], multiferroics [14], superconductor [6], metal-insulator transitions [15] , and other correlated systems [7]. In analogy, microwave spectroscopy could also provide critical insights into material systems with novel physics that are harder to experimentally access, such as thin films with flake sizes[1]restricted to about 10 µm; an example of particular interest to our group is analog quantum simulators (AQS) with spins in semiconductors [3].

There are several existing techniques for GHz broadband spectroscopy. Please see Table III for comparison of the previous techniques to our technique.

This paper presents a new technique for performing broadband microwave spectroscopy on flakes with lateral sizes as small as one µm (could be even smaller using lithography with finer resolution) and nm-sized thicknesses. The crucial aspect is a coplanar waveguide (CPW) with a series gap in the signal line over the flake. This series gap concentrates the sensitivity of the macroscopic device into the area of the flake; as a result, the macroscopic transmission in a transmission line (of length meters, going through various magnetic field, temperature, etc.) is dominated by the ac conductance of the flake. After introducing the technique, we present simulations for a candidate system (twisted bilayer graphene), and thus demonstrate a simulated ability to deduce the ac conductance (including interesting frequency dependence due to the superconducting gap) while including real-world irreproducibility.

## III. The Proposed Technique

| | Definition | Stage of Experiment |
|---|---|---|

[1] we note that, throughout this paper, our use of the word "flake" includes samples which are not exfoliated but which, because of other fabrication limitations, cannot be made larger than 10 µm in lateral size

| | | |
|---|---|---|
| $Z_{CPW}$ | Characteristic impedance | |
| $Z(\omega)$ | Sample impedance (Fig. 2) | |
| $t(\omega)$ | Transmission (Eq. 3) | |
| $V_1$ | Amplitude of AC voltage | |
| $Z_{gap}$ | impedance from the combination of $C_{Couple}$, $C_{Series}$, and $Z(\omega)$ | |
| $C_{Series}$ | Series capacitance (Fig. 2) | |
| $C_{Couple}$ | Coupling capacitance (Fig. 2) | |
| $C_{SH}$ | Shunt capacitance (Fig. A1) | |
| $t_{sim}^{Z}(\omega)$ | Simulated transmission with material $Z(\omega)$ and non-ideal wiring | Simulated Measurement |
| $t_{sim}^{\infty}(\omega)$ | Simulated transmission with insulator ( $Z(\omega)=\infty$) and non-ideal wiring | Simulated Control |
| $t_{wir\text{-}a}(\omega)$, $t_{wir\text{-}b}(\omega)$ | Measured transmission in wiring | |
| $\alpha$ | Constant near unity | |
| $\sigma_1$, $\sigma_2$ | Real and imaginary parts of sample 2-D conductivity (units sq/$\Omega$) | |
| $t_{\infty}$ | Similar to $t_{sim}^{\infty}(\omega)$, but with ideal wiring | |
| $t_R(\omega)$ | Sample transmission relative to control | Both |
| $t_{exp}^{flake}(\omega)$ | Experimental transmission with material $Z(\omega)$ and non-ideal wiring | Experimental Measurement |
| $t_{exp}^{control}(\omega)$ | Experimental transmission with sample in insulating regime and non-ideal wiring | Experimental Control |

*Table 1 Definitions of symbols. "Stage of experiment" refers to either i) "Measurement" with sample* $Z(\omega)$ *or ii) "Control" with* $Z(\omega) = \infty$.

The basic idea (see Fig. 1) is to fabricate a CPW with a series gap in the signal line over the flake of interest. The crucial item here is that the gap in the signal line means that transmission is dominated by the conductivity of the thin-film flake; because of this, we will show that modest changes in the conductivity of the thin film effect large changes in the transmitted signal. This technique will allow groups to sensitively and easily measure conductivity spectra for a wide variety of materials without requiring obscure or specialized fabrication or measurement techniques. Once fabricated, one would conveniently measure the transmission characteristic using standard Vector Network Analyzer methods.

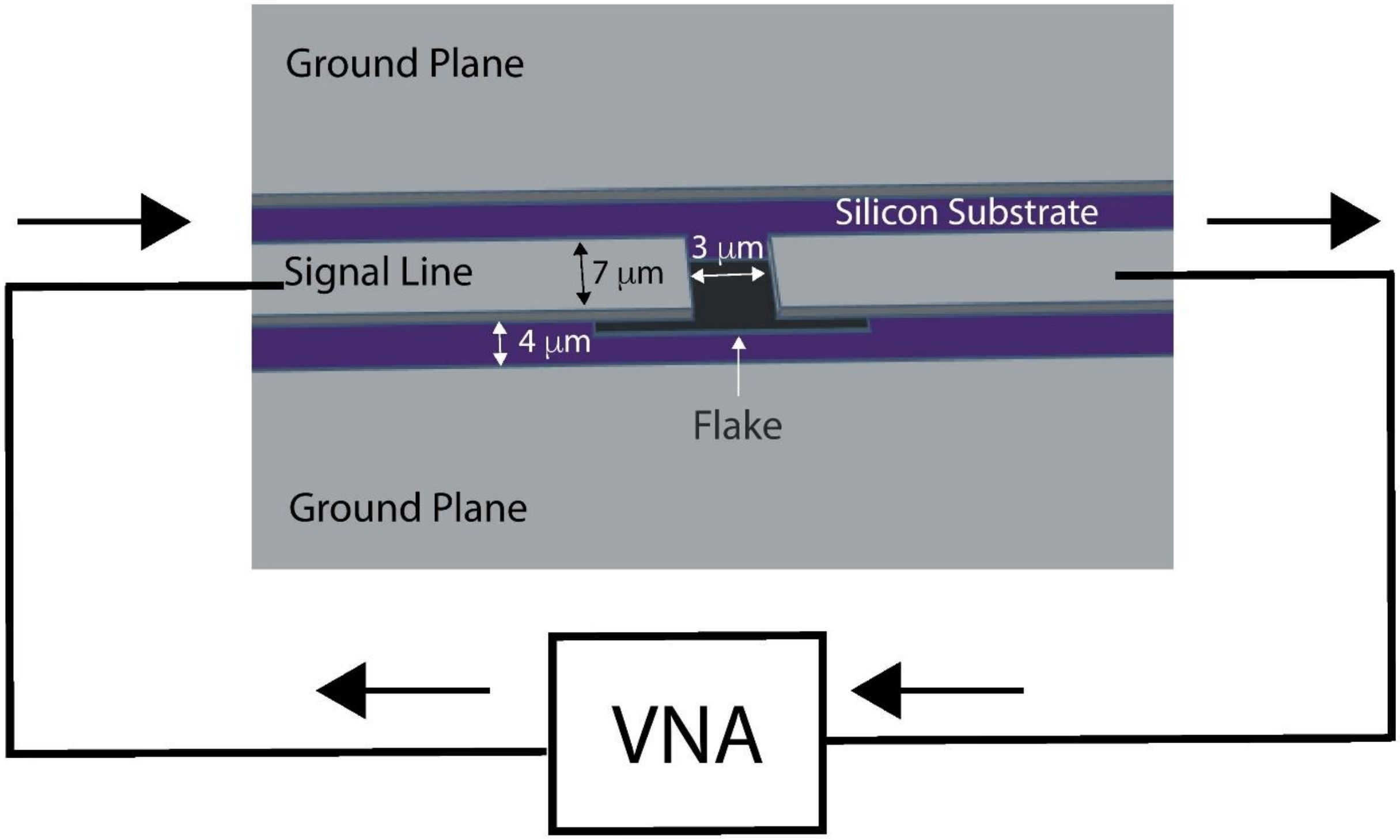

Figure 1 The proposed experimental setup for the technique. The gap in the CPW (grey ground plane and grey signal line) signal line would be located directly over the thin-film flake (black), which would be separated from the CPW by a dielectric; VNA is a network analyzer. Dimensions are illustrative, chosen to ensure ease of fabrication (requiring only photo-lithography as opposed to techniques such as electron-beam lithography) and impedance matching with 50 Ω cables. The crucial theme: the gap in the signal line means that transmission is dominated by the conductivity of the thin-film flake.

The equivalent circuit is shown in Fig. 2, including the characteristic impedance $Z_{CPW}$. As mentioned above, we emphasize that the desired large signal arises in the parameter regime where conduction from left to right is dominated by $1/Z(\omega)$ , the ac conductance of the flake of interest. In Appendix A, we give a full derivation of Equations, and also include a discussion of the fact that the change in geometry near the gap does not significantly perturb the current flow nor the value of transmission t(ω).We now present a brief derivation of the relationship between the t(ω) and the thin film's ac conductance. This derivation just uses continuity of current, and voltage drops across impedances. To start, assuming the gap is much smaller than the wavelength of the microwave signal, continuity of current yields [in this derivation, we do not carry along the explicit frequency dependence of t(ω)]:

$$1 = r + t \qquad \text{Equation (1)}$$

Similarly, combining this with voltage drops yields:

$$V_1(1 + r) = V_1\, t + \frac{V_1}{Z_{CPW}} t Z_{gap} = V_1\, t + \frac{\frac{V_1}{Z_{CPW}}}{i\omega C_{Series} + \frac{1}{\frac{2}{i\omega C_{Couple}} + Z(\omega)}} t \qquad \text{Equation (2)}$$

Here, $V_1$ is the amplitude of the incoming signal at frequency ω, $r$ and $t$ are the reflection and transmission coefficients, respectively, and $Z_{gap}$ is the impedance from the combination of $C_{couple}$, $C_{series}$ and Z(ω). If the system were resistively coupled to the signal lines instead, we would replace $\frac{1}{i\omega C_{Couple}}$ by a contact resistance in Equation (2). Combining the above equations yields:

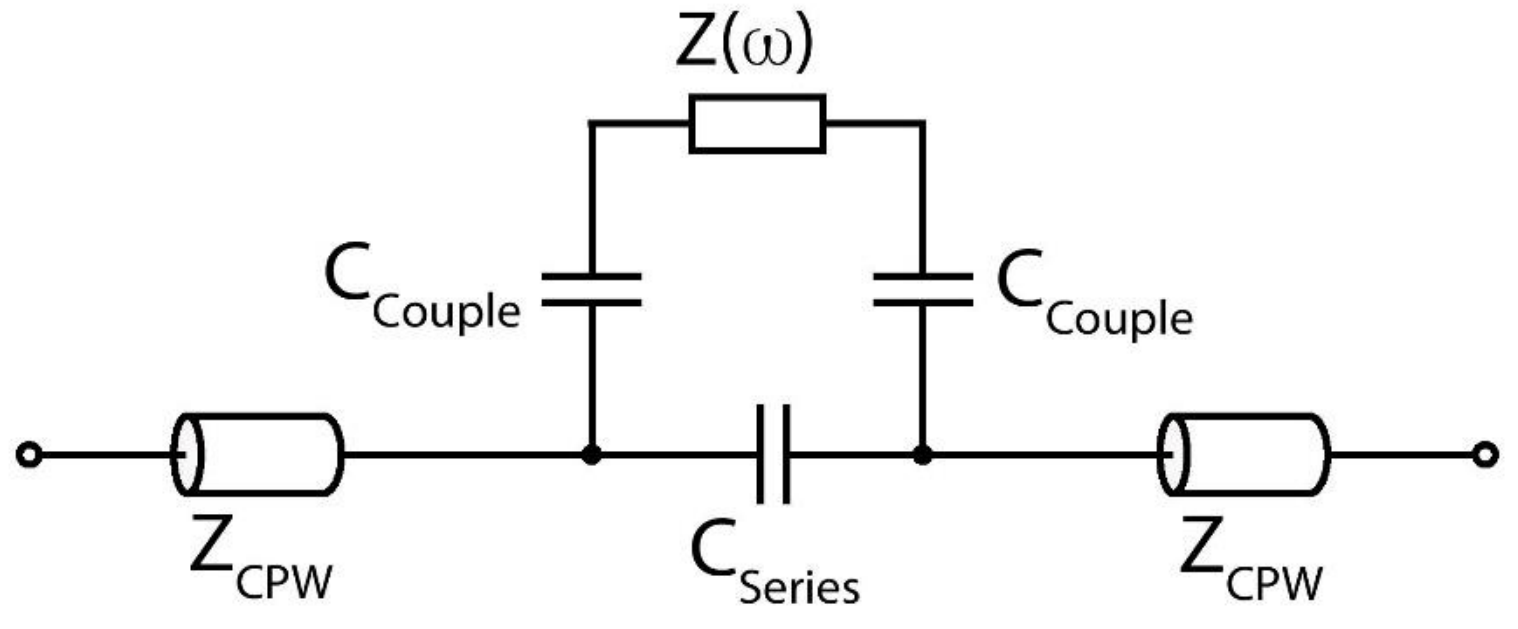

Figure 2 Circuit diagram. $Z_{CPW}$ is the characteristic impedance of the transmission line, $C_{Series}$ is the direct capacitance (in the absence of the conducting flake) across the gap, $C_{Couple}$ is the (vertical) capacitance due to the overlap between the signal line and the flake, Z(ω) is the complex conductance of the flake; typically, $C_{Couple} \gg C_{Series}$

$$t(\omega) = \frac{2\left[\frac{1}{\frac{2}{i\omega C_{Couple}}+Z(\omega)}+i\omega C_{Series}\right]}{2\left[\frac{1}{\frac{2}{i\omega C_{Couple}}+Z(\omega)}+i\omega C_{Series}\right]+\frac{1}{Z_{CPW}}} \qquad \text{Equation (3)}$$

Finally, a VNA (vector network analyzer) typically reports magnitude and phase of $t(\omega)$.

We can note some simple, appealing limits for Equation (3): i) In the limit of a continuous (vanishingly small gap) CPW (where the series capacitance dominates the conduction, and where $|1/(i\omega C_{Series})| \ll Z_{CPW}$, $t(\omega) \cong 1$; this is appealing because it shows that a continuous CPW has ideal transmission. ii) in the opposite limit where the conduction is dominated by the flake $|1/(i\omega C_{Series})| \gg |Z(\omega)| \gg |1/(i\omega C_{Couple})|, Z_{CPW}$, we obtain $t(\omega) \cong (2Z_{CPW})/Z(\omega) \ll 1$, showing that the transmission and its frequency dependence allow deduction of the flake ac conductance. We note that, for the parameters in Table II, all three conditions in limit ii) can be satisfied for $|Z(\omega)|$ varying by a range of a factor of 25; for instance, at 1 GHz, all three conditions are satisfied if $|Z(\omega)| \in [7,150]\,k\Omega$ (please see Fig. 4 for a graphical elucidation of parameter range).

| Parameter | Value or limit | Source | Z ($2\pi\ 10^8$ Hz) |
|---|---|---|---|
| $C_{Series}$ | 500 aF | Simulation (see Appendix) | $3 \times 10^6\ \Omega$ |
| $C_{Couple}$ | 50 fF | Simulation (see Appendix) | $3 \times 10^4\ \Omega$ |
| $C_{SH}$ (see Fig. A1 and discussion in Appendix) | Less than 1 fF | Bounds on simulation (see Appendix) | greater than $2 \times 10^6\ \Omega$ |
| Flake size | 7 μm X 7 μm | [1] | |
| Insulating layer thickness | 30 nm | | |

Table II: Values or bound on useful parameters, obtained using the dimensions in Figure 1, including the approximate impedance at 100 MHz. Please see the Appendix for details.

**Simulated Spectral Measurements**

We now discuss simulations for broadband measurements of superconducting twisted bilayer graphene encapsulated in h-BN on a Si substrate [1] with the setup depicted in Figure 1 (the graphene is represented by the black rectangle). We chose this material because i) the superconducting transition temperature of about 0.5 K means that there is interesting frequency dependence near 10 GHz, ii) the microwave spectrum has not been previously measured, and iii) the superconducting and insulating states of such sample offer a convenient in-situ way to perform a control experiment, i.e., to measure with non-zero flake conductance and then with essentially zero flake conductance (these two stages are referred to in Table I as “Measurement” and “Control”). However, as we discuss later, we note that the typical superconducting complex conductivity (pure imaginary below the gap frequency) leads to some complications which are not present in most materials of interest.

We will now show simulations, with the ultimate goal of demonstrating the ability to deduce, from a simple GHz transmission measurement, i) qualitative spectral features such as resonant states in an AQS structure or a superconducting gap and ii) quantitative information of the complex conductivity of a $\mu$m-scale flake (Figure 3 (d)) while including real-world non-idealities in wiring. We also show the large difference between the situations of flake conductivity present versus absent (insets in Figures 3 (b) and (c)); we will refer to the flake absent (i.e., zero conductivity) as the “control” experiment. The large difference demonstrates both i) the inherent ability to observe resonant features such as those in AQS samples, and ii) the large signal-to-noise ratio inherent in our technique.

The proposed experiment depends on taking transmission spectra of the twisted bilayer graphene with small $|Z(\omega)|$, and then large $|Z(\omega)|$ as a control measurement (i.e., under zero magnetic field and then under a magnetic field large enough to make the twisted bilayer graphene non-superconducting).

Figure 3 shows simulations using our proposed technique to characterize superconducting twisted bilayer graphene, using the dimensions in Figure 1; *please note that the dimensions are illustrative; our technique is applicable over a wide range of sizes*. We assume appropriate sizes and thicknesses as in Table II; amongst other conditions, this gives a large enough $C_{couple}$ to allow us to satisfy conditions ii) above.

We start with the real and imaginary parts of the complex two-dimensional conductivity spectra ($\sigma_1$ and $\sigma_2$, respectively, with units of sq/$\Omega$) in Fig. 3(a), as calculated using the Mattis-Bardeen theory [16] for an ideal superconductor. The calculation is based on superconducting gap energy $2\Delta=3.5k_BT_C\cong0.09$ meV for $T_C$ = 300 mK, and non-superconducting (pure real) conductivity $\sigma_n=4\times10^{-4}\ \Omega^{-1}$; we obtained both values from transport measurements taken on twisted bilayer graphene with a twist angle of $\mathbf{1.16}°$ [1]. We note that the conductivity has the classic features for a superconductor: i) $\sigma_2\rightarrow\infty$at low frequency, and ii) the dissipation $1/\sigma_1$ becomes finite above the frequency $\Delta/h$. Please see Appendix for details.

We then simulate the measured complex transmission $t_{sim}^{Z}(\omega)$ and $t_{sim}^{\infty}(\omega)$, for the conducting and insulating cases [Table I "Measurement" and "Control] (Figures 3 (b) and (c) show the complex magnitude and phase). In these two graphs, the insets show the "raw" simulations, and the main parts show the normalized ratio $t_{sim}^{Z}(\omega) / t_{sim}^{\propto}(\omega)$. These two simulations allow us to take into account the effects of non-ideal wiring, and using the ratio allows us to minimize those effects. Taking into account non-ideal wiring through two different wiring transmission functions $t_{wir\text{-}a}(\omega)$ and $t_{wir\text{-}b}(\omega)$ (please see Appendix for details), we have used

$Z(\omega)=\alpha/(\sigma_1-i\sigma_2)$

$$t_{sim}^{Z}(\omega)=t(\omega)\, t_{wir\text{-}a}(\omega) \quad \text{and} \quad t_{sim}^{\infty}(\omega)=t_{\infty}(\omega)\, t_{wir\text{-}b}(\omega)$$

where α is a constant of order one to estimate the number of squares in the flake (we use 3 μm/7 μm), t is defined by Eq'n 3 using the parameters in Table II, and $t_\infty=t|_{(Z(\omega)=\infty)}$. In this way, in the main parts of Figures 3b and 3c we have used the protocol of flake stage and then control stage to minimize the effects of non-ideal wiring:

$$t_R(\omega) \equiv \frac{t_{sim}^{Z}(\omega)}{t_{sim}^{\infty}(\omega)} = \frac{t(\omega)}{t_\infty}\left[\frac{t_{wir-a}(\omega)}{t_{wir-b}(\omega)}\right] \qquad \text{Equation (4)}$$

where we note that using the ratio ($t_{wir\text{-}a}(\omega)$ / $t_{wir\text{-}b}(\omega)$) allows us to remove the effects of most of the non-ideal wiring transmission and has only the irreproducible parts of the transmission.

We note that Figures 3(b) and 3(c) demonstrate clearly the two claims made at the beginning of this section: Transmission measurements will allow us to immediately observe resonant features, and there is a large difference in magnitude and phase between the measurement with the flake and without (conductivity turned off). In particular, please note the easily-detectible resonance at about 20 GHz in Fig. 3c, due to the superconducting transition; it is this type of frequency feature for which we have devised our technique. We also note, as mentioned earlier, that the atypical complex conductivity of a superconductor means that magnitude of t does not show the spectral feature [the peak at about 14 GHz in Fig. 3b is an artifact arising from the pure imaginary value of Z(w), which leads to the series impedance of $C_{couple}$ and Z(ω) going through zero]; this is because near 20 GHz, |σ| is dominated by $\sigma_2$ which does not exhibit a large spectral feature at the gap, and thus $t \cong 2\, Z_{CPW} / Z(\omega)$ does not exhibit a large spectral feature. In contrast, we have verified (not shown) that for a more typical pure real conductivity (for example, a single Drude resonance), both |t(ω)| and arg(t(ω)) show the spectral feature of the resonance.

We now go beyond demonstrating the ability to qualitatively see spectral structure, and demonstrate how accurately we can deduce the complex conductivity of a μm-scale flake. In Figure 3(d) we have simulated reverting the simulated transmission data t(ω), which incorporates real-world non-idealities ($t_{wir\text{-}a}(\omega)$ and $t_{wir\text{-}b}$), to recover quantitative complex conductivity data (please see Appendix for details). In the following prescription, we explain both i) how we simulated results for Figure 3(d), and ii) how we propose to use a parallel prescription for experimental data (for which one would start with data equivalent to that in Figures 3b and 3c).

We start with the ratio defined in Equation 4, and extend it to include either the simulation or the proposed protocol in the actual experiment:

$t_R$ (ω)≡ $t_{sim}^Z$ (ω)/($t_{sim}^\infty$ (ω) )=t(ω)/$t_\infty$ [$t_{wir\text{-}a}$(ω) /$t_{wir\text{-}b}$(ω)] OR $t_R$(ω)=$t_{exp}^{flake}$ (ω)/$t_{exp}^{control}$ (ω)

Next, we calculate the effective

$t_{eff}$ (ω) ≡ $t_R$ (ω) $t_\infty$=t(ω)[($t_{wir\text{-}a}$(ω))/($t_{wir\text{-}b}$(ω))] OR $t_{eff}$ (ω)≡ $t_R$ (ω) $t_\infty$=$t_{exp}^{flake}$(ω)/($t_{exp}^{control}$ (ω)) $t_\infty$

where the first possibility was used to generate Figure 3(d), and in the second possibility we would use the conducting and control data in the experiment, and where the parameter values in $t_\infty = t|_{Z(\omega)=\infty}$ came from Table II for the simulations in Figure 3(d), but for the experiment would come from either i) estimates using device geometry or ii) as deduced from low-frequency data (please see Appendix). Finally, we have

$$\sigma(\omega) = \alpha\left[\left(\frac{t_R(\omega)}{2} Z_{CPW}\big(1 - t_R(\omega)\big) - i\omega C_{Series}\right)^{-1} - \frac{2}{i\omega C_{Couple}}\right]^{-1} \qquad \text{Equation (5)}$$

Using this prescription, we note in Figure 3d the excellent sensitivity, and the ability to deduce the true complex conductivity [note the excellent agreement between Figures 3 (a) and 3 (d)]. We note that we lose sensitivity to Z(ω) at low frequencies, where $C_{couple}$ dominates. Finally, for quantitative comparison to the "Shunt in CPW" technique in Table III, please see Appendix B for simulation results with the same flake dimensions. In this case, the analogous signal sizes are about 1 dB and π/20.

We can now consider the optimum regime for our technique. In order for this prescription to work, we need the impedance in the gap of the CPW to be dominated by the flake impedance (condition ii above); this sets limits on the flake impedance which are functions of frequency. In Fig. 4, we show these frequency-dependent limits for the parameters in Table II.

Finally, having shown the qualitative and quantitative advantages and limitations of our proposed technique, we can now show in a condensed fashion the comparison with previous techniques – please see Table III.

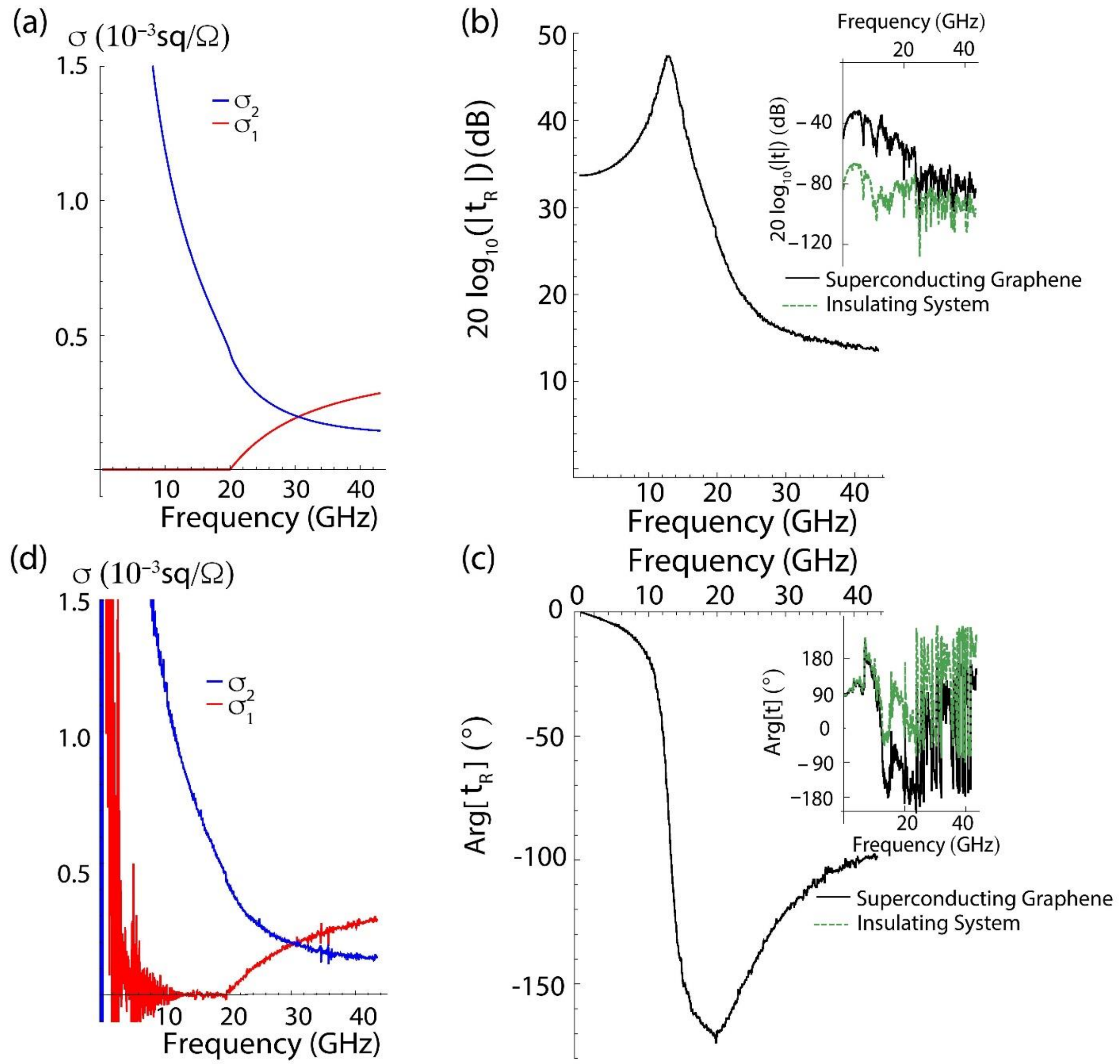

Figure 3: Expected transmission signal for superconducting twisted bilayer graphene, and reversion to obtain complex conductivity. a) Complex conductivity obtained from Mattis-Bardeen theory (see Appendix); b) magnitude and c) phase (main part) predicted normalized transmission signals – note that the spectral features arising from the superconducting gap would be represented as a large feature in the phase, demonstrating the qualitative strength of the proposed technique. Please also note that these simulations include effects of non-ideal wiring. b) magnitude and c) phase (insets) show the un-normalized spectra, where the effects of non-ideal wiring are manifestly obvious. d) Demonstration of ability to revert data back to complex

conductivity, using the prescription in Equations 5 in the main text; the excellent agreement between panels a) and d) demonstrate the quantitative strength of the proposed technique.

| Technique | Advantage(s) | Disadvantage(s) | Reference(s) |
|---|---|---|---|
| Free-space | No fabrication | Can't measure μm-scale | [14] |
| Shunt in CPW | | Can't measure μm-scale; may require insulator on top of film. | [10] |
| AC Transport | Convenient fabrication | Can't measure μm-scale nor above 5 MHz | [17] |
| Dielectric properties of biological cells | Convenient fabrication. | Was only applied to measuring frequency-independent R and C, not Z(ω); requires insulator on top of film. | [18] |
| AFM μwave spectroscopy | Good spatial resolution and signal sensitivity (about 1 dB); can measure μm-scale flakes. | Requires AFM. Quite slow due to need to move between each frequency. | [19] |
| Gapped CPW (present technique) | Convenient fabrication; wide range of thin film resistivities; can get quantitative frequency dependence with excellent sensitivity (1 dB or less) | Requires insulator on top of film; requires location with respect to flake(s). | |

Table III: Comparison of different broadband microwave techniques.

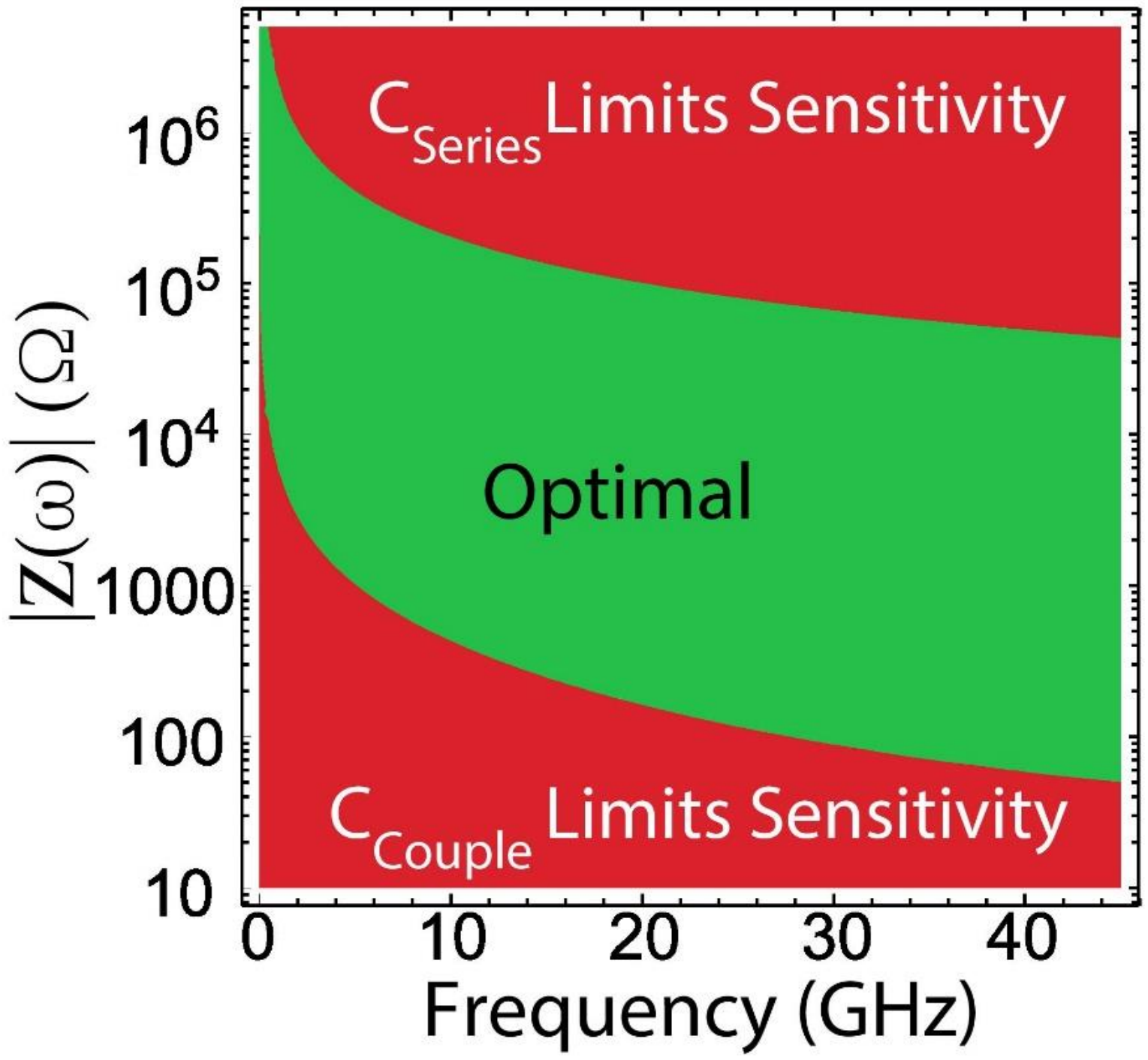

Figure 4: Phase space of high (green) and low (red) sensitivity using conditions ii) discussed above. As discussed in the text, the range of sensitive measurements would be extended by at least a factor of ten in $|Z(\omega)|$ if we use fabrication techniques with better resolution than optical lithography.

## IV. Experimental considerations

One could improve the sensitivity of the new measurement technique to less conducting and/or smaller flakes (well below a few μm) by using fabrication techniques with better spatial resolution. In particular, since in general the sensitivity is limited by the condition $|Z(\omega)| << |1/(i\omega C_{series})|$, and since the value of $C_{series}$ is roughly proportional to the width of the CPW, fabricating a device whose width and gap are n times smaller can increase the maximum detectable resistivity by n. We could easily achieve a factor of ten larger range of $|Z(\omega)|$ by using electron-beam lithography for the central portion of the series-gapped CPW.

In addition, as mentioned above, making reliable Ohmic contact to 2-D materials with lateral sizes of order mm is often challenging [20]; we note that it is even more challenging to make such contacts with high-quality transmission line characteristics (e.g., avoiding large impedance mismatch in vias). In contrast, with our proposed technique, we have obviated the need for such contacts.

Challenges are also present: The first is driven by the need to have a control (i.e., insulating) measurement for comparison. We have chosen graphene as a candidate material in part because the superconducting transition provides a very convenient control. However, for twisted bilayer graphene in particular, we may need to implement a buried gate to achieve superconductivity [1], or we might apply gate voltages on the CPW centerline. For other candidate materials, it may be necessary to run a control experiment with a separate chip in the same cryostat cooldown, or perhaps even in different cooldowns; we note that in the simulations in Fig. 3, we used an irreproducibility in the wiring transmission which is appropriate for a single cooldown with controllable $Z(\omega)$, but probably smaller than that which we could expect for different cooldowns.

In addition, we note that in some cases the finite real conductivity $\sigma_1$ will be a complicating factor. In particular, the effective RC delay in the bottom plate of the capacitor $C_{couple}$ could possibly distort the reversion to $\sigma(\omega)$. We have analyzed the case of a distributed RC network [21], and found that the distortion is minimal (of order 10 %) in the optimum range of Figure 4. We plan on presenting the details of this in a future publication.

Also, we note that the magnitude of the total transmission (including real-world insertion loss) as shown in Fig. 3b is quite small. We note that i) much of the insertion loss occurs in the wiring before the sample, so that a larger input power will help; ii) it may be necessary to use low-temperature amplifiers (a common technique).

Another challenge is presented by the sparsity of flake coverage in many two-dimensional materials (often of order 1 % relative areal density). In such case, one will need to take care to either/both i) orient the CPW and gap with respect to predetermined flake locations or ii) fabricate many gapped CPW's in order to have a good statistical chance to overlap flake(s). We note that choice ii) would also naturally provide the desired control device on the same chip.

Finally, we hope to extend the usefulness of this technique by making wafer-scale buried gapped CPW's before deposition/exfoliation/deposition of the flakes of interest, a la' the work of [1].

## V. Conclusions

We have proposed a technique for performing sensitive broadband microwave conductivity spectroscopy on materials only available as micron-scale flakes, using a coplanar waveguide with a series gap above the material of interest. We have shown through theory and simulations that the conductivity of the flake effects large changes in the amplitude and phase of the transmitted signal, and can allow us to qualitatively measure important spectral features. By including real-world non-idealities, we have also shown that the new technique can be used to accurately measure broadband spectral conductivity, even if the spectral reproducibility of measurements is limited. We note the relative advantages and disadvantages summarized in Table III.

More broadly, this setup could be used to measure the conductivity spectrum of many two-dimensional materials of interest, even when the lateral size is limited to μm scale, ranging from metallic systems to pinned modes of charge density waves to AQS structures [1, 9]. These results are also striking for the range of resistivity for which the transmitted signal is sensitive, which is large enough to characterize such diverse systems as pinning modes from insulating electron solids [5, 9] and the free carrier response in highly conducting systems like twisted bilayer graphene.

## VI. Acknowledgements

The authors would like to offer special thanks to Josh Pomeroy of NIST for numerous insightful conversations and logistical and technical assistance with the low-temperature transmission measurements. We would also like to thank Zac Barcikowski, Richard Silver, Pradeep Namboodiri, Ranjit Kashid, Richard Silver, Nikolai Zhitenev, and Michael Stewart (all of NIST) for insightful conversations, and for technical training and assistance throughout this process.

## AUTHOR DECLARATIONS

### Conflict of Interest

The authors declare no conflict of interest.

### DATA AVAILABILITY

Data sharing is not applicable to this article as no new data were created or analyzed in this study.

**Appendix A: Derivations of Equations in the Main Text**

This appendix provides a derivation of Equations 1-3 in the main text; we simply use continuity of current, and voltage drops across impedances. We define current polarity as positive when moving to the right.

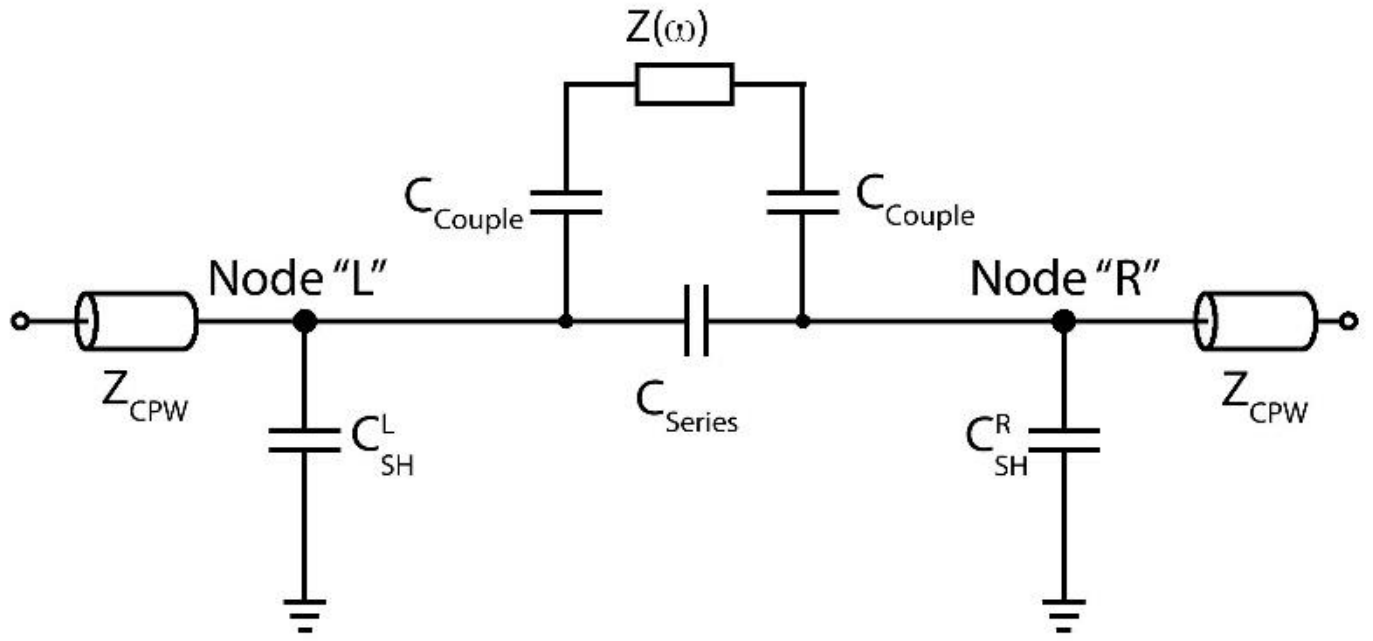

Figure A1: The same circuit as Figure 1, with the addition of the explicit $Z_{CPW}$ and nodes "L" and "R".

1) Perfect transmission lines and definition of r and t

We start by considering the portions of Figure A1 that correspond to perfect transmission lines (node "L" and everything to the left of it, node "R" and everything to the right of it); anywhere in this area, by definition of the characteristic impedance $Z_{CPW}$, we have $I = \pm V / Z_{CPW}$, where the negative sign corresponds to waves moving to the left (reflected). The signal to the left of node "L" consists of an incident and reflected wave. The signal to the right of node "R" consists entirely of a transmitted wave. The relationships between the voltages and currents of these signals are governed by the telegrapher's equations (current continuity and voltage drops across impedances).

We now consider the reflection and transmission coefficients $r$ and $t$, respectively. At node "L", the incoming, reflected and transmitted voltages are given by $V_1\, e^{i(\omega t - kz)}$ , $rV_1\, e^{i(\omega t + kz)}$, and $tV_1\, e^{i(\omega t - kz)}$; here, $V_1$ is the amplitude of the incoming wave, $z$ is the position along the circuit, and $k=\omega/c\, \surd(\varepsilon_{Eff}\, \mu_{Eff})$, where $\varepsilon_{Eff}$ ($\mu_{Eff}$) is the effective relative permittivity (magnetic permeability) of the CPW. From this point on, since the wavelength at GHz frequencies is much larger than the gap size, all voltages are given by amplitudes only. Thus, the total voltage at node "L" is $(1 + r)\, V_1$. In addition, using the definition of characteristic impedance $Z_{CPW}$, given above, the current at node "L" (in particular, in the horizontal branch to the left of "L") is $(1-r)\, V_1/Z_{CPW}$. Similarly, at node "R", the voltage and current are given by $tV_1$ and $tV_1/Z_{CPW}$, respectively.

2) Consideration of $C_{SH}$

We now consider $C_{SH}$; The presence of the gap perturbs the relationship between I and V from that of the characteristic impedance, due to electric field lines that no longer propagate at right angles between the signal line and the ground of the CPW; we will show in this Appendix that, for the frequency and impedance regime of interest (the green area in Figure 4), this perturbation has a negligibly small effect.

The presence of non-zero $C_{SH}$ results in current flow through the impedance $Z_{gap}$ as follows:

Node "L": $I_{gap} = (1 - r)\, V_1 / Z_{CPW} - (1 + r)\, V_1\, (i\, \omega\, C^L_{SH})$ Equation (A1)

Node "R": $I_{gap} = t\, V_1 / Z_{CPW} + t\, V_1\, (i\, \omega\, C^R_{SH})$ Equation (A2)

Here, $I_{gap}$ is the current flowing in the horizontal branch to the right of "L" and to the left of "R"; also, $Z_{gap}$ is the series-parallel combination of $C_{couple}$, $C_{series}$, and $Z(\omega)$, which from the values in Table II and Figure 4 is less than about 1 M$\Omega$.

Note that, from the values in Table II, at any frequency below 100 GHz, $1 / Z_{CPW} >> \omega\, C^L_{SH}$, $\omega\, C^R_{SH}$ and thus we can immediately see from Eq'n A2 that $C^R_{SH}$ is negligible. It turns out that, at node "L" to first order in $\omega\, C^L_{SH}\, Z_{CPW}$, $C^L_{SH}$ is also negligible. A simple way to see that is to observe that to first order, the non-reflected part of the wave $(1 – r) \cong 2\, \omega\, C^L_{SH}\, Z_{CPW} + 2\, Z_{CPW} / Z_{gap}$ gets larger as $C^L_{SH}$, and thus the loss of current running through $C^L_{SH}$ (the negative second term in Eq. A1) is exactly balanced by an increase in the non-reflected current at "L"; finally, the first-order estimate ($\omega\, C_{SH}\, Z_{CPW} << 1$) yields $t \cong 2\, Z_{CPW} / Z_{gap}$, which is the same estimate derived from Eq'n 3 in the main text. For the rest of this Appendix, and in the main text, we thus suppress any further discussion of $C_{SH}$.

3) Telegrapher equations (current continuity and voltage drops)

Using repeatedly the suppression of $C_{SH}$, current continuity requires that the current entering node "L" be equal to the current exiting node "R":

$$(1-r)\frac{V_1}{Z_{CPW}} = \frac{tV_1}{Z_{CPW}} \qquad \text{Equation (A3)}$$

This is equivalent to Equation (1) in the main text.

Considering voltage drops, we have

$V_L = V_R + I_{gap}\, Z_{gap}$,

where $V_L$ and $V_R$ are the voltages at the two nodes and with $1 / Z_{gap} = 1 / (2 / i\, \omega\, C_{couple} + Z(\omega)) + i\, \omega\, C_{series}$. Substituting from the above, we have

$(1+r)=t+(t/Z_{CPW})Z_{gap}$ Equation (A4)

This is Equation (2) in the main text, after substituting for $Z_{gap}$. Finally, by combining current continuity and voltage drops, we can solve for either r or t:

$$t = 2 / (2 + Z_{gap} / Z_{CPW}) = (2 / Z_{gap}) / (2 / Z_{gap} + 1 / Z_{CPW}) \qquad \text{Equation (A5)}$$

This is equivalent to Equation (3) in the main text.

**Appendix B: Transmission for Technique Entitled "Shunt in CPW"**

Similar to Appendix A, in this Appendix we present the transmission for a previous technique: a flake below a CPW without any gap, to provide a comparison to the new technique.

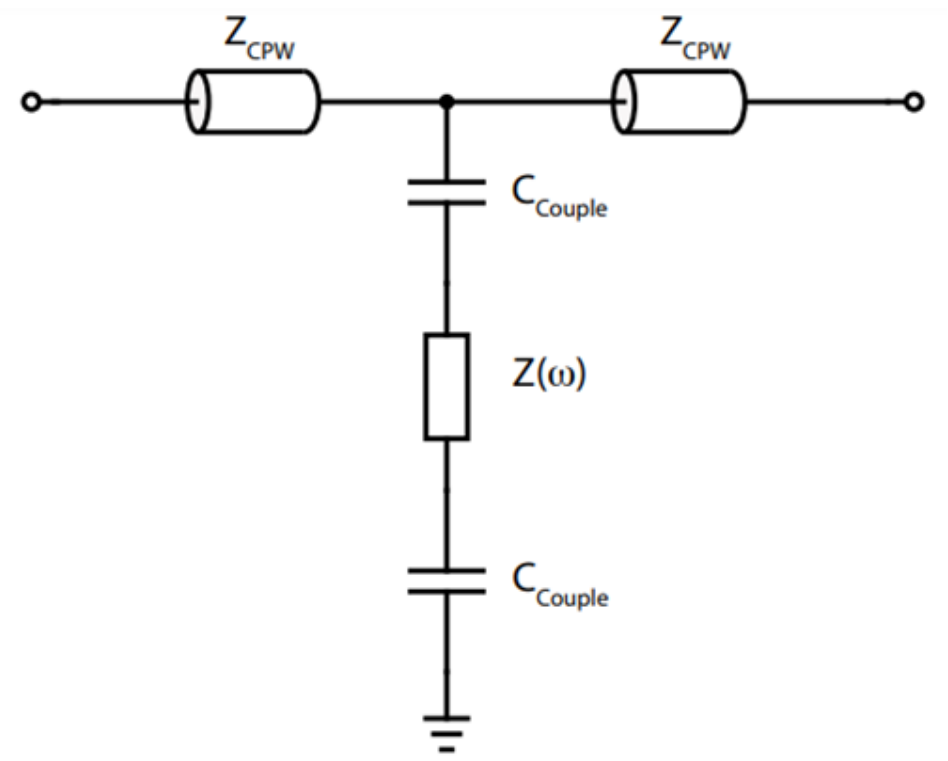

Figure B1: circuit diagram for "Shunt in CPW" technique

Very briefly, by following the same procedure as in Appendix A, we obtain from current continuity

$$1/Z_{CPW} - r\, 1/Z_{CPW} = t\, 1/Z_{CPW} + t/(Z(\omega)+2/(i\omega C_{Couple})) \qquad \text{Equation (B1)}$$

and from voltage drops

$$V_1 + rV_1 = tV_1 \qquad \text{Equation (B2)}$$

Combining Equations (B1) and (B2):

$$t = 2/(2+Z_{CPW}/(Z(\omega)+2/(i\omega C_{Couple}))) \qquad \text{Equation (B3)}$$

From this, we can see very clearly the crucial difference: since $C_{couple}$ is now in parallel with $Z_{CPW}$ (rather than in series as in the main text and Appendix A), and for a small flake with lateral area of order (1 $\mu$m)$^2$, we have 1 /($\omega C_{couple}$)>~ 2 k$\Omega$ >> $Z_{CPW}$. Thus, for this “shunt in CPW” technique, the flake has very little effect on the transmission (t $\cong$ 1). We can also see this in Fig B2, where the signal sizes are much smaller than in Figure 4.

For comparison, we can show the simulated (ideal wiring) sensitivity achievable with the “shunt in CPW” for a wide range of flake conductances (Figure B2 a and b) and with the gapped CPW (c and d). Note the much larger response than for the previous shunt technique.

## Appendix C: Simulations of Values/Uncertainties for Capacitances; how to deduce two capacitances from data; results of deduced complex conductivity using uncertainties

### Part I: Simulations of Capacitance Values

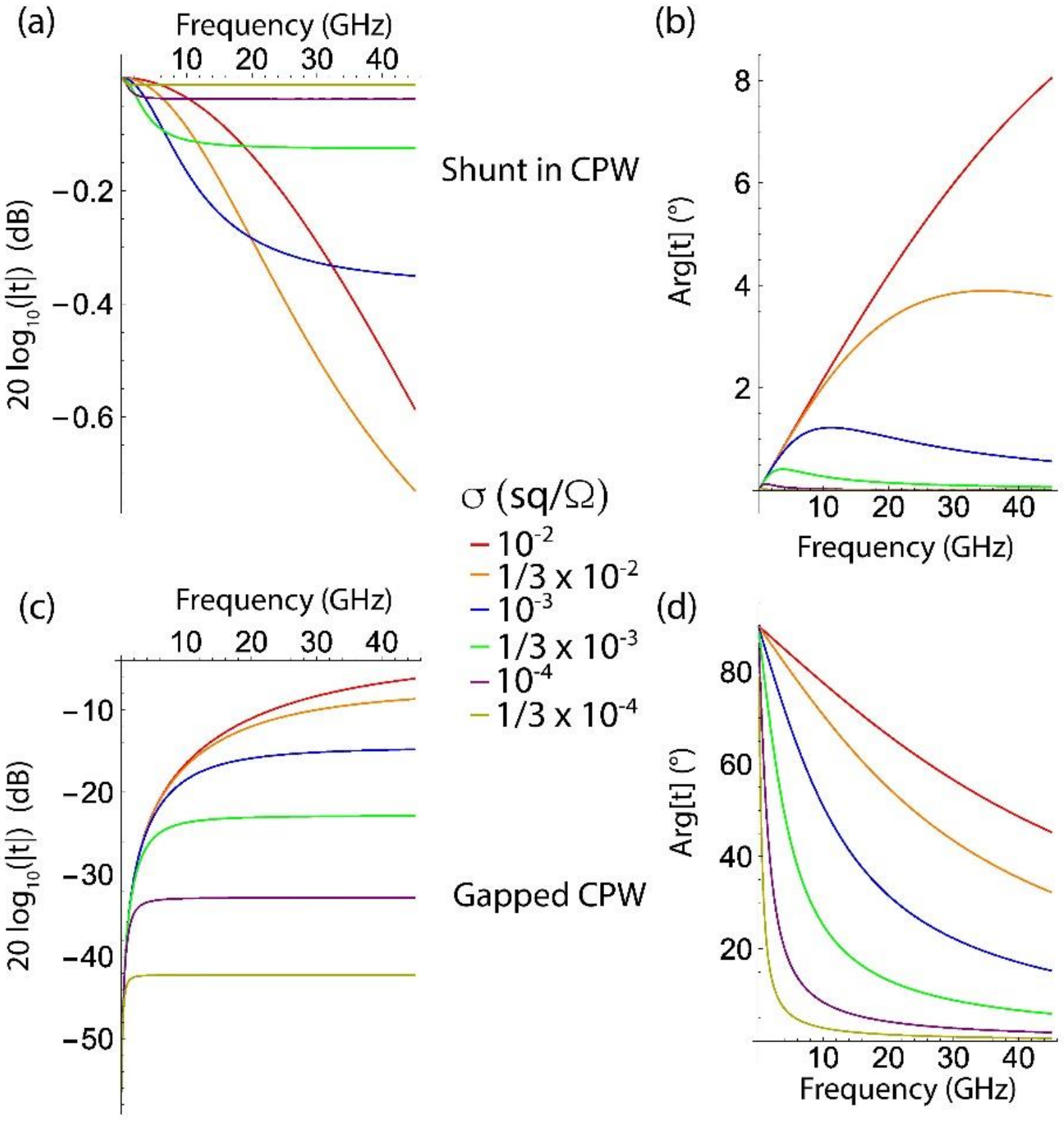

We simulated capacitance values for the structure in Figure 1 using COMSOL [22] with $C_{series}$ calculated in the absence of a flake beneath the CPW gap. The hardest parameter to deduce is $C_{SH}$, because it represents the modification in the capacitance between signal line and ground due to the removal of the metal in the gap. In order to put an upper bound on this parameter, we simulated the capacitance between signal line and ground plane in two structures with length 250 μm: i) A continuous CPW, and ii) a CPW with a gap of length 3 μm; all other dimensions were the same as in Figure 1. We then made an upper bound for $C_{SH}$ using two simple approaches: i) With an approximate capacitance per unit length of 45 fF / 250 μm, we obtain an approximate capacitance for signal line of length 3 μm of about 0.54 fF; ii) the values obtained (of order 40 fF) differed by slightly less than 0.5 fF. Thus, we have used a conservative upper bound of 1 fF in Table II and in the main text.

**Part II: Deduction of $C_{couple}$ and $C_{series}$ from geometry and from experimental data, and Uncertainties**

As mentioned in the main text, we could simply estimate $C_{couple}$ and $C_{series}$ from the geometry;

Fig. B2 Amplitude (a) and phase (b) of transmission spectra for for 2D 7 μm×7 μm flakes at a variety of 2D resistivities using "shunt in CPW" method. Amplitude (c) and phase (d) of transmission spectra for the configuration shown in Fig. 1 (gapped CPW). All parameters are from Table I. Note the much larger sizes of signals in the gapped CPW technique;

as shown in Part III of this Appendix, a 10 % uncertainty in this estimate corresponds to about a 10 % uncertainty in $\sigma$.

We can also estimate $C_{couple}$ and $C_{series}$ from the experimental data; the starting point is to assume that we are able to do the experiments, i.e., that we are in the high-sensitivity (green) parameter range in Figure 4.

**$C_{series}$:** We use the conditions corresponding to the control experiment, where the flake is insulating or absent, so that the transmission is dominated by $C_{series}$: $Z(\omega) >> 1/(\omega C_{series}) >> Z_{CPW}$; this condition holds for the frequency range 100 MHz × (3 MΩ / $Z(\omega)$) << f << 100 GHz. Using Eq'n 3 and including the measured non-idealities $t_{wir}(\omega)$ in the wiring, we can derive in this limit that

$t(\omega) \cong 2\, i\, \omega\, C_{series}\, Z_{CPW}\, t_{wir}(\omega)$ Equation (C1)

Thus, to deduce $C_{series}$, we propose to i) verify that $t(\omega) / t_{wir}(\omega) \propto \omega$, and then ii) use the slope and the measured non-ideality $t_{wir}(\omega)$ (total insertion loss) to get the value of $C_{series}$,.

**$C_{couple}$:** We use the high-conducting state of the flake at low frequency, so that the impedance of $C_{couple}$ dominates the transmission: $2 / \omega\, C_{couple} >> Z(\omega)$, $Z_{CPW}$, along with the assumption that $C_{couple} >> C_{series}$. This is, in essence, in the lower red region of Fig. 4. The first assumption sets a frequency limit of f << 100 MHz (60 kΩ / $Z(\omega)$) using the values in Table II.

Again from Eq'n 3 and including the non-idealities, it is straightforward to derive

$t(\omega) = i\, \omega\, C_{couple}\, Z_{CPW}\, t_{wir}(\omega)$, Equation (C2)

and thus we can easily deduce $C_{couple}$; we propose to i) verify that $t(\omega) / t_{wir}(\omega) \propto \omega$, and then ii) use the slope and the measured non-ideality $t_{wir}(\omega)$ (total insertion loss) to get the value of $C_{couple}$.

**Part III: Estimation of Uncertainties in experimental values of $C_{couple}$ and $C_{series}$**

Using propagation of errors from Equations C1 and C2, we can derive, e.g.,

$\Delta C_{couple} / C_{couple} = [(\Delta t / t)^2 + (\Delta \omega / \omega)^2 + (\Delta Z_{CPW} / Z_{CPW})^2 + (\Delta t_{wir}(\omega) / t_{wir}(\omega))]^2]^{1/2}$. In this value, we estimate that the dominant contribution is from the statistical uncertainty in comparing $t_{wir}(\omega)$ from two different measurements; from Figure D1, we can see that this uncertainty in the non-ideality of the wiring is less than 10 %. Thus, our estimate in the relative uncertainties of the experimentally deduced values of both $C_{couple}$ and $C_{series}$ are 10 %.

**Part IV: Effect of uncertainties in deduced capacitance values on complex conductivity**

As stated in the previous section, we estimate the relative uncertainty in the deduced values of $C_{Series}$ and $C_{Couple}$ to be 10%. To explore the effects of uncertainty in $C_{Couple}$ and $C_{Series}$ on the deduced $\sigma(\omega)$, we took the simulated experimental results in Figs. 3b and 3c, and deduced $\sigma(\omega)$ values with Equations 3 and 10, using $C_{Series}$ and $C_{Couple}$ that deviated from their "actual" values by ± 10%. The results are shown in Fig. D1 alongside the "correct" values, shown in the dashed black curves. Simply put, the relative uncertainties in the deduced conductivity values are the same as the relative uncertainties in the capacitance values.

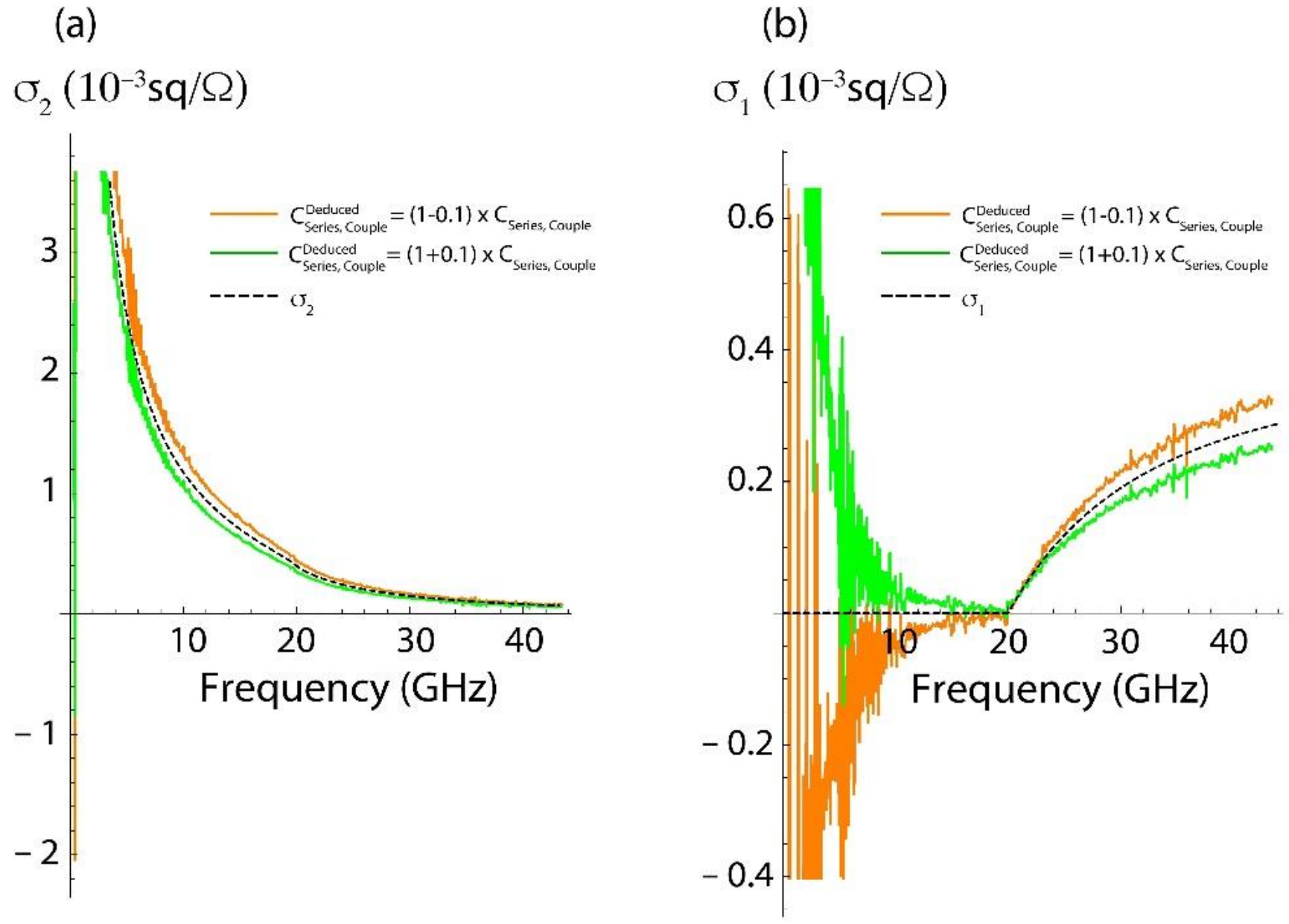

**Figure C1:** Deduced values of imaginary and real parts of complex conductivity, taking into account 10 % uncertainties in the deduced capacitances. Note that the relative deviation in $\sigma(\omega)$ is the same (10 %).

**Appendix D: Details of Simulations in Figure 3**

**Part I: Non-ideal Insertion Loss in Wiring**

In order to have a realistic simulation of the ability to deduce the complex, frequency-dependent $Z(\omega)$ in Figure 3, we need to include the non-ideal insertion loss in the wiring; this includes both overall nonzero insertion loss, low-pass filtering due to cable resistance and capacitance, and in addition the inevitable resonances. Figure D1 shows the transmission for two repeated measurements, where the difference is in disconnecting and reconnecting the room-temperature cable assemblies. These measurements were taken on a continuous (not gapped) CPW on a Si substrate, with wire bonds from contact pads on the Si to a PCB. The PCB was mounted in a 4 K cryostat. The total length of wiring both inside and outside the cryostat was about 6 m, with two connections. We can see that the irreproducibility is about 0.5 dB and a few degrees, which is sufficiently small to allow the measurements outlined in the main text.

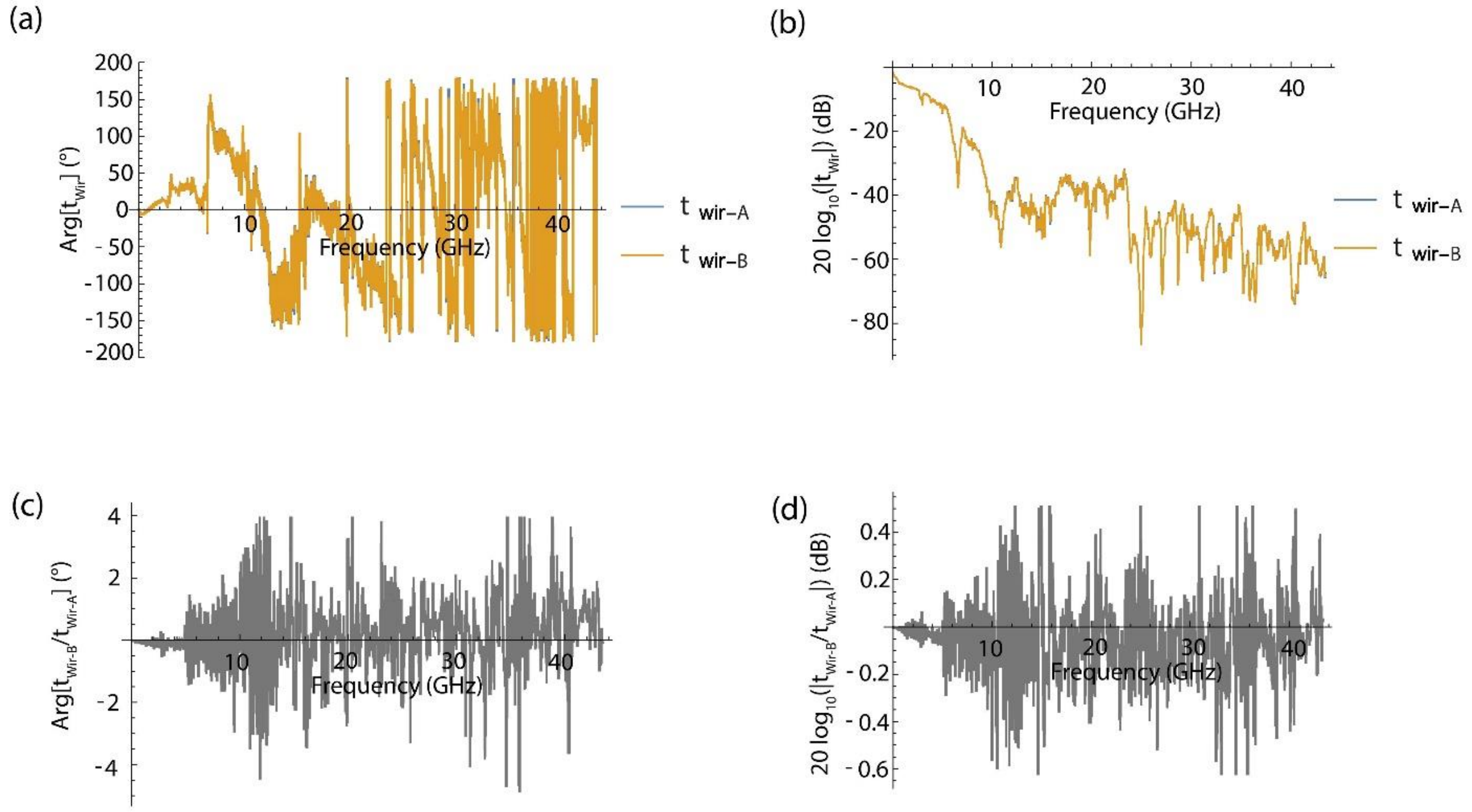

Fig. D1 The phase (a) and amplitude (b) of transmission spectra measured for a continuous CPW on a Si substrate placed in a dry $^4$He cryostat, “A” and “B” denote measurements before and after the connections between the VNA ports and the ports leading to the cryostat were disconnected and reconnected. The difference in phase and amplitude of the transmission spectra are shown in (c) and (d), respectively.

**Part II: Mattis-Bardeen formulae for $\sigma_1$ and $\sigma_2$**

As mentioned in the main text, the conductivity spectrum of the superconducting state of twisted bilayer graphene was simulated using the values from Ref. [1] in the Mattis-Bardeen equation [16].

$$\frac{\sigma_1(\omega)}{\sigma_N} = 2\int_{\epsilon_0}^{\infty} \frac{[f(E)-f(E+\hbar\omega)]g(E)}{\hbar\omega} dE + \int_{\epsilon_0-\hbar\omega}^{-\epsilon_0} \frac{[1-2f(E+\hbar\omega)]g(E)}{\hbar\omega} dE \qquad \text{Equation (E1)}$$

$$\frac{\sigma_2\,(\omega)}{\sigma_N} = \frac{1}{\hbar\omega}\int_{Max(\epsilon_0-\hbar\omega,-\epsilon_0)}^{\epsilon_0} \frac{[1-2f(E+\hbar\omega)]\left(E^2+\epsilon_0^2+\hbar\omega E\right)}{\left[\epsilon_0^2-E_0^2\right]^{\frac{1}{2}}\left[(E+\hbar\omega)^2-\epsilon_0^2\right]^{\frac{1}{2}}}dE \qquad \text{Equation (E2)}$$

where $g(E) = \frac{E^2+\epsilon_0^2+\hbar\omega E}{\left[E_0^2-\epsilon_0^2\right]^{\frac{1}{2}}\left[(E+\hbar\omega)^2-\epsilon_0^2\right]^{\frac{1}{2}}}$, $\epsilon_0$is half the energy of the superconducting gap, $\sigma_N$ is the conductivity of the normal metallic state, and f(x) is the Fermi-Dirac distribution function. Note that these equations use the convention μ=0.